\documentclass[aps,preprint]{revtex4}
\usepackage{amssymb,epsf}
\usepackage{amsmath}
\usepackage{graphicx}
\begin{document}
\title{Lorentzian Wormholes in Lovelock Gravity}
\author{M. H. Dehghani$^{1,2}$\footnote{email address:
mhd@shirazu.ac.ir} and Z. Dayyani$^{1}$}
\affiliation{$^1$ Physics Department and Biruni Observatory, College of Sciences, Shiraz University, Shiraz 71454, Iran\\
        $^2$ Research Institute for Astrophysics and Astronomy of Maragha (RIAAM), Maragha,
        Iran}

\begin{abstract}
In this paper, we introduce the $n$-dimensional Lorentzian wormhole solutions of third
order Lovelock gravity. In contrast to Einstein gravity and as
in the case of Gauss-Bonnet gravity,
we find that the wormhole throat radius, $r_0$, has a lower limit that depends
on the Lovelock coefficients, the dimensionality of the spacetime and
the shape function.
We study the conditions of having normal matter near the
throat, and find that the matter near the throat can be normal
for the region $r_0 \leq r \leq r_{\max}$, where $r_{\max}$ depends on
the Lovelock coefficients and the shape function. We also
find that the third order Lovelock term with negative
coupling constant enlarges the radius of the region of normal matter, and
conclude that the higher order Lovelock terms with negative
coupling constants enlarge the region of normal matter near the
throat.
\end{abstract}

\maketitle
\section{ Introduction}

Wormholes are tunnels in the geometry of space and time that connect two
separate and distinct regions of spacetimes. Although such objects were long
known to be solutions of Einstein equation, a renaissance in the study of
wormholes has taken place during 80's motivated by the possibility of quick
interstellar travel \cite{Mor1}. Wormhole physics is a specific example of
adopting the reverse philosophy of solving the gravitational field equation,
by first constructing the spacetime metric, then deducing the stress-energy
tensor components. Thus, it was found that these traversable wormholes
possess a stress-energy tensor that violates the standard energy conditions
(see, e.g., \cite{Mor2}, \cite{Vis} or \cite{Lobo} for a more recent review).
The literature is rather extensive
in candidates for wormhole spacetimes in Einstein gravity, and one may
mention several cases, ranging from wormhole solutions in the presence of
the cosmological constant \cite{Mann}, wormhole geometries in higher
dimensions \cite{high-dim}, to geometries in the context of linear and
nonlinear electrodynamics \cite{Electro}. Also the stability of wormhole
solutions has been analyzed by considering specific equations of state \cite
{stability1}, or by applying a linearized radial perturbation around a
stable solution \cite{stability2}.

One of the main areas in wormhole research is to try to avoid, as much as
possible, the violation of the standard energy conditions. For static
wormholes of Einstein gravity the null energy condition is violated, and
thus, several attempts have been made to somehow overcome this problem. In
order to do this, some authors resort to the alternative theories of gravity:
the wormhole geometries of Brans-Dicke theory have been investigated in \cite
{Brans}; of Kaluza-Klein theory in \cite{Kaluz}; and of a higher curvature
gravity in \cite{HDgr}. In the latter, it was found that the weak energy
condition may be respected in the throat vicinity of the wormholes of higher
curvature gravity. A special branch of higher curvature gravity which
respects the assumptions of Einstein --that the left-hand side of the field
equations is the most general symmetric conserved tensor containing no more
than two derivatives of the metric-- is the Lovelock gravity \cite{Lov}. This
theory represents a very interesting scenario to study how the physics of
gravity are corrected at short distance due to the presence of higher order
curvature terms in the action. Static solutions of second and third orders
Lovelock gravity have been introduced in \cite{Boul} and \cite{Deh},
respectively. For wormholes with small throat radius, the curvature near the
throat is very large, and therefore the investigation of the effects of
higher curvature terms becomes important. The possibility of obtaining
a wormhole solution from the instanton solutions of Lovelock gravity
has been studied in \cite{Mar}. The wormhole solutions of dimensionally
continued Lovelock
gravity have been introduced in \cite{Li}, while these kind of solutions in second
order Lovelock gravity and the possibility of obtaining solutions with
normal and exotic matter limited to the vicinity of the throat have been
explored in \cite{Kar}. Here, we want to add the third order term of
Lovelock theory to the gravitational field equations, and investigate the
effects of it on the possibility of having wormhole solutions with normal
matter. We also want to explore the effects of higher order Lovelock terms
on the region of normal matter near the throat.

The outline of this paper is as follows. We give a brief review of the field
equations of third order Lovelock gravity and introduce the wormhole
solutions of this theory in Sec. \ref{Stat}. In Sec. \ref{Exotic}, we present
the conditions of having normal matter near the throat and exotic matter
everywhere. We finish our paper with some
concluding remarks.

\section{Static Wormhole solutions\label{Stat}}

We, first, give a brief review of the field equations of third order Lovelock
gravity, and then we consider the static wormhole solutions of the theory.
The most fundamental assumption in standard general relativity is the
requirement that the field equations be generally covariant and contain at
most second order derivative of the metric. Based on this principle, the
most general classical theory of gravitation in $n$ dimensions is the
Lovelock gravity. The Lovelock equation up to third order terms without
the cosmological constant term may be written as \cite{Hois}
\begin{equation}
G_{\mu \nu }^{(1)}+\sum_{p=2}^{3}\alpha _{i}^{\prime
}\left( H_{\mu \nu }^{(p)}-\frac{1}{2}g_{\mu \nu }\mathcal{L}^{(p)}\right)
=\kappa _{n}^{2}T_{\mu \nu },  \label{FF}
\end{equation}
where $\alpha _{p}^{\prime }$'s are Lovelock coefficients, $T_{\mu \nu }$ is
the energy-momentum tensor, $G_{\mu \nu }^{(1)}$ is just the Einstein
tensor, $\mathcal{L}^{(2)}=R_{\mu \nu \gamma \delta }R^{\mu \nu \gamma
\delta }-4R_{\mu \nu }R^{\mu \nu }+R^{2}$ is the Gauss-Bonnet Lagrangian,
\begin{eqnarray}
\mathcal{L}^{(3)} &=&2R^{\mu \nu \sigma \kappa }R_{\sigma \kappa \rho \tau
}R_{\phantom{\rho \tau }{\mu \nu }}^{\rho \tau }+8R_{\phantom{\mu
\nu}{\sigma \rho}}^{\mu \nu }R_{\phantom {\sigma \kappa} {\nu \tau}}^{\sigma
\kappa }R_{\phantom{\rho \tau}{ \mu \kappa}}^{\rho \tau }+24R^{\mu \nu
\sigma \kappa }R_{\sigma \kappa \nu \rho }R_{\phantom{\rho}{\mu}}^{\rho }
\notag \\
&&+3RR^{\mu \nu \sigma \kappa }R_{\sigma \kappa \mu \nu }+24R^{\mu \nu
\sigma \kappa }R_{\sigma \mu }R_{\kappa \nu }+16R^{\mu \nu }R_{\nu \sigma
}R_{\phantom{\sigma}{\mu}}^{\sigma }-12RR^{\mu \nu }R_{\mu \nu }+R^{3}
\label{L3}
\end{eqnarray}
is the third order Lovelock Lagrangian, and $H_{\mu \nu }^{(2)}$ and $H_{\mu
\nu }^{(3)}$ are
\begin{equation}
H_{\mu \nu }^{(2)}=2(R_{\mu \sigma \kappa \tau }R_{\nu }^{\phantom{\nu}%
\sigma \kappa \tau }-2R_{\mu \rho \nu \sigma }R^{\rho \sigma }-2R_{\mu
\sigma }R_{\phantom{\sigma}\nu }^{\sigma }+RR_{\mu \nu }),  \label{Love2}
\end{equation}
\begin{eqnarray}
H_{\mu \nu }^{(3)} &=&-3(4R^{\tau \rho \sigma \kappa }R_{\sigma \kappa
\lambda \rho }R_{\phantom{\lambda }{\nu \tau \mu}}^{\lambda }-8R_{%
\phantom{\tau \rho}{\lambda \sigma}}^{\tau \rho }R_{\phantom{\sigma
\kappa}{\tau \mu}}^{\sigma \kappa }R_{\phantom{\lambda }{\nu \rho \kappa}%
}^{\lambda }+2R_{\nu }^{\phantom{\nu}{\tau \sigma \kappa}}R_{\sigma \kappa
\lambda \rho }R_{\phantom{\lambda \rho}{\tau \mu}}^{\lambda \rho }  \notag \\
&&-R^{\tau \rho \sigma \kappa }R_{\sigma \kappa \tau \rho }R_{\nu \mu }+8R_{%
\phantom{\tau}{\nu \sigma \rho}}^{\tau }R_{\phantom{\sigma \kappa}{\tau \mu}%
}^{\sigma \kappa }R_{\phantom{\rho}\kappa }^{\rho }+8R_{\phantom
{\sigma}{\nu \tau \kappa}}^{\sigma }R_{\phantom {\tau \rho}{\sigma \mu}%
}^{\tau \rho }R_{\phantom{\kappa}{\rho}}^{\kappa }  \notag \\
&&+4R_{\nu }^{\phantom{\nu}{\tau \sigma \kappa}}R_{\sigma \kappa \mu \rho
}R_{\phantom{\rho}{\tau}}^{\rho }-4R_{\nu }^{\phantom{\nu}{\tau \sigma
\kappa }}R_{\sigma \kappa \tau \rho }R_{\phantom{\rho}{\mu}}^{\rho
}+4R^{\tau \rho \sigma \kappa }R_{\sigma \kappa \tau \mu }R_{\nu \rho
}+2RR_{\nu }^{\phantom{\nu}{\kappa \tau \rho}}R_{\tau \rho \kappa \mu }
\notag \\
&&+8R_{\phantom{\tau}{\nu \mu \rho }}^{\tau }R_{\phantom{\rho}{\sigma}%
}^{\rho }R_{\phantom{\sigma}{\tau}}^{\sigma }-8R_{\phantom{\sigma}{\nu \tau
\rho }}^{\sigma }R_{\phantom{\tau}{\sigma}}^{\tau }R_{\mu }^{\rho }-8R_{%
\phantom{\tau }{\sigma \mu}}^{\tau \rho }R_{\phantom{\sigma}{\tau }}^{\sigma
}R_{\nu \rho }  \notag \\
&&-4RR_{\phantom{\tau}{\nu \mu \rho }}^{\tau }R_{\phantom{\rho}\tau }^{\rho
}+4R^{\tau \rho }R_{\rho \tau }R_{\nu \mu }-8R_{\phantom{\tau}{\nu}}^{\tau
}R_{\tau \rho }R_{\phantom{\rho}{\mu}}^{\rho }+4RR_{\nu \rho }R_{%
\phantom{\rho}{\mu }}^{\rho }-R^{2}R_{\mu \nu }),  \label{Love3}
\end{eqnarray}
respectively.

As in the paper of Morris and Thorne \cite{Mor1}, we adopt the reverse
philosophy in solving the third order Lovelock field equation, namely, we
first consider an interesting and exotic spacetime metric, then finds the
matter source responsible for the respective geometry. The generalized
metric of Morris and Thorne in $n$ dimensions may be written as
\begin{equation}
ds^{2}=-e^{2\phi (r)}dt^{2}+\left( 1-\frac{b(r)}{r}\right)
^{-1}dr^{2}+r^{2}d\theta
_{1}^{2}+\sum\limits_{i=2}^{n-2}\prod\limits_{j=1}^{i-1}\sin ^{2}\theta
_{j}d\theta _{i}^{2},  \label{met1}
\end{equation}
where $\phi (r)$ and $b(r)$ are the redshift function and shape function,
respectively. Although the metric coefficient $g_{rr}$ becomes divergent at
the throat of the wormhole $r=r_{0}$, where $b(r_{0})=r_{0}$, the proper
radial distance
\begin{equation*}
l(r)=\int_{r_{0}}^{r}\frac{dr}{\sqrt{1-b/r}}
\end{equation*}
is required to be finite everywhere. The metric (\ref{met1}) represents a
traversable wormhole provided the function $\phi (r)$ is finite everywhere
and the shape function $b(r)$ satisfies the following two conditions:
\begin{eqnarray}
1)\text{ }b(r) &\leq &r,  \label{Cond1} \\
2)\text{ \ }rb^{\prime } &<&b,  \label{Cond2}
\end{eqnarray}
where the prime denotes the derivative with respect to $r$. The first
condition is due to the fact that the proper radial distance should be real
and finite for $r>r_{0}$, and the second condition comes from the
flaring-out condition \cite{Mor1}.

The mathematical analysis and the physical interpretation will be simplified
using a set of orthonormal basis vectors
\begin{eqnarray*}
\mathbf{e}_{\hat{{t}}} &=&e^{-\phi }\frac{\partial }{\partial t},\text{
\ \ \ }\mathbf{e}_{\hat{r}}=\left( 1-\frac{b(r)}{r}\right) ^{1/2}\frac{%
\partial }{\partial r}, \\
\mathbf{e}_{\hat{1}} &=&r^{-1}\frac{\partial }{\partial \theta _{1}},%
\text{ \ }\mathbf{e}_{\hat{i}}=\left( r\prod\limits_{j=1}^{i-1}\sin
\theta _{j}\right) ^{-1}\frac{\partial }{\partial \theta _{i}}. \label{Basis}
\end{eqnarray*}
Using the orthonormal basis (\ref{Basis}), the components of
energy-momentum tensor $T_{_{\hat{\mu
}\hat{\nu }}}$ carry a simple physical interpretation, i.e.,
\begin{equation*}
T_{_{\hat{t}\hat{t}}}=\rho ,\text{ \ \ }T_{_{\hat{r}\hat{r}%
}}=-\tau ,\text{ \ \ }T_{_{\hat{i}\hat{i}}}=p,
\end{equation*}
in which $\rho (r)$ is the energy density, $\tau (r)$ is the radial tension,
and $p(r)$ is the pressure measured in the tangential directions orthogonal
to the radial direction. The radial tension $\tau (r)=-p_{r}(r)$, where
$p_{r}(r)$ is the radial pressure. Using a unit system
with $\kappa_n^2=1$, and defining
$\alpha _{2}\equiv (n-3)(n-4)\alpha _{2}^{\prime }$ and $\alpha
_{3}\equiv (n-3)...(n-6)\alpha _{3}^{\prime }$ for simplicity, the nonvanishing
components of Eq. (\ref{FF}) reduce to
\begin{eqnarray}
\rho (r) &=&\frac{(n-2)}{2r^{2}}\Big\{-\left( 1+\frac{2\alpha
_{2}b}{r^{3}}+\frac{3\alpha _{3}b^{2}}{r^{6}}\right) \frac{(b-rb^{\prime })}{%
r}  \notag \\
&&\quad \quad +\frac{b}{r}\left[ (n-3)+(n-5)\frac{\alpha _{2}b}{r^{3}}+(n-7)%
\frac{\alpha _{3}b^{2}}{r^{6}}\right] \Big\},  \label{rho} \\
\tau (r) &=&\frac{(n-2)}{2r}\Big\{-2\left( 1-\frac{b}{r}%
\right) \left( 1+\frac{2\alpha _{2}b}{r^{3}}+\frac{3\alpha _{3}b^{2}}{r^{6}}%
\right) \phi ^{\prime } \notag \\
&&\quad \quad +\frac{b}{r^{2}}\left[ (n-3)+(n-5)\frac{\alpha _{2}b}{r^{3}}%
+(n-7)\frac{\alpha _{3}b^{2}}{r^{6}}\right] \Big\},  \label{tau}
\end{eqnarray}
\begin{eqnarray}
p(r) &=&\left( 1-\frac{b}{r}\right)
\left( 1+\frac{2\alpha _{2}b}{r^{3}}+\frac{3\alpha _{3}b^{2}}{r^{6}}\right) %
\left[ \phi ^{\prime \prime }+{\phi ^{\prime }}^{2}+\frac{(b-rb^{\prime
})\phi ^{\prime }}{2r(r-b)}\right] \   \notag \\
&&+\left( 1-\frac{b}{r}\right) \left( \frac{\phi ^{^{\prime }}}{r}+\frac{%
b-b^{^{\prime }}r}{2r^{2}(r-b)}\right) \left[ (n-3)+(n-5)\frac{2\alpha _{2}b%
}{r^{3}}+(n-7)\frac{3\alpha _{3}b^{2}}{r^{6}}\right]   \notag \\
&&-\frac{b}{2r^{3}}\left[ \left( n-3\right) \left( n-4\right) +\left(
n-5\right) \left( n-6\right) \frac{\alpha _{2}b}{r^{3}}+\left( n-7\right)
\left( n-8\right) \frac{\alpha _{3}b^{2}}{r^{6}}\right]   \notag \\
&&-\frac{2\phi ^{^{\prime }}}{r^{4}}\left( 1-\frac{b}{r}\right) \left(
b-b^{^{\prime }}r\right) \left( \alpha _{2}+3\alpha _{3}\frac{b}{r^{3}}%
\right).  \label{pr}
\end{eqnarray}

\section{Exoticity of the matter \label{Exotic}}

To gain some insight into the matter threading the wormhole, one should
consider the sign of $\rho$, $\rho -\tau $ and $\rho+p$.
If the values of these functions are nonnegative, the weak energy
condition (WEC) ($T_{\mu \nu }u^{\mu }u^{\nu }\geq 0$, where $u^{\mu }$ is
the timelike velocity of the observer) is satisfied, and therefore the
matter is normal. In the case of negative $\rho $, $\rho -\tau $ or $\rho+p$, the WEC
is violated and the matter is exotic. We consider a specific class of
particularly simple solutions corresponding to the choice of $\phi (r)=$
\textrm{const.}, which can be set equal to zero without loss of generality.
In this case, $\rho -\tau $ and $\rho+p$ reduce to
\begin{eqnarray}
\rho -\tau &=&-\frac{(n-2)}{2r^{3}}\left( b-rb^{\prime
}\right)\left( 1+\frac{2\alpha _{2}b%
}{r^{3}}+\frac{3\alpha _{3}b^{2}}{r^{6}}\right),\label{exo}\\
\rho+p&=&-\frac{\left( b-rb^{\prime
}\right)}{2r^{3}}\left( 1+\frac{6\alpha _{2}b%
}{r^{3}}+\frac{15\alpha _{3}b^{2}}{r^{6}}\right)\nonumber\\
&& +\frac{b}{r^3}\left\{ (n-3)+(n-5)\frac{2\alpha _{2}b}{r^{3}}%
+(n-7)\frac{3\alpha _{3}b^{2}}{r^{6}}\right\}.
\label{Exo}
\end{eqnarray}
\subsection{Positivity of $\rho$ and $\rho+p$}
Here, we investigate the conditions of positivity of $\rho $ and
$\rho+p$ for different choices of shape function $b(r)$.
\subsubsection{Power law shape function:}
First, we consider the positivity of $\rho $ and $\rho+p$ for the power law shape function $%
b=r_{0}^{m}/r^{m-1}$ with positive $m$. The positivity of $m$ comes from the
conditions (\ref{Cond1}) and (\ref{Cond2}). The functions $\rho $ and $\rho+p$ for the power law
shape function are positive for $r>r_{0}$ provided $r_{0}>r_{c}$, where $%
r_{c}$ is the largest positive real root of the following equations:
\begin{eqnarray}
&& (n-3-m)r_{c}^{4}+(n-5-2m)\alpha _{2}r_{c}^{2}+(n-7-3m)\alpha _{3}=0,
\nonumber\\
&& (2n-6-m)r_{c}^{4}+2\alpha _{2}(2n-10-3m)r_{c}^{2}+3\alpha _{3}(2n-14-5m)=0. \label{rcpow}
\end{eqnarray}
Of course if Eqs. (\ref{rcpow}) have no real root, then there is
no lower limit for $r_0$ and $\rho $ and $\rho+p$ are positive everywhere.

\subsubsection{Logarithmic shape function:}

Next, we investigate the positivity of $\rho $ and $\rho+p$ for logarithmic shape function, $%
b(r)=r\ln r_{0}/\ln r$. In this case the conditions (\ref{Cond1}) and (%
\ref{Cond2}) include $r_{0}>1$. The functions $\rho $ and $\rho+p$
are positive for $r>r_{0}$ provided $r_{0}\geq r_{c}$, where $r_{c}$ is
the largest real root of the following equations:
\begin{eqnarray}
&& \left[(n-3)r_c^4+\alpha_2(n-5) r_c^2+\alpha_3(n-7)\right]\ln r_{c}-(r_{c}^{4}+2\alpha _{2}r_{c}^2+3\alpha_3)=0,\nonumber\\
&& 2\left[(n-3)r_c^4+2\alpha_2(n-5) r_c^2+3\alpha_3(n-7)\right]\ln r_{c}-(r_{c}^{4}+6\alpha _{2}r_{c}^2+15\alpha_3)=0.
\label{rclog}
\end{eqnarray}
If $r_c>1$, then $\rho $ and $\rho+p$  are positive for $r>r_0 \geq r_c$, but
in the case that Eqs. (\ref{rclog}) have no real positive
root or their real roots are less than $1$, then the lower limit for
$r_0$ is  just $1$, and $\rho $ and $\rho+p$  are positive for $r \geq r_0>1$

\subsubsection{Hyperbolic solution:}

Finally, we consider the positivity of density $\rho $ and $\rho+p$  for the hyperbolic
shape function, $b(r)=r_{0}\tanh (r)/\tanh (r_{0})$ with $r_{0}>0$, which
satisfies the conditions (\ref{Cond1}) and (\ref{Cond2}). The
functions $\rho $ and $\rho+p$  will be positive provided $r_{0}>r_{c}$, where $r_{c}$ is the
largest real root of the following equations:
\begin{eqnarray}
&& \left( n-4\right) r_{c}^{4}+\alpha _{2}\left( n-7\right) r_{c}^{2}+\alpha
_{3}\left( n-10\right) ++\frac{r_{c}^{5}+2\alpha_2r_{c}^{3}+3\alpha_3r_{c}}{\sinh r_c \cosh r_c} =0,\nonumber\\
&& \left(2n-7\right) r_{c}^{4}+2\alpha _{2}\left(2n-13\right) r_{c}^{2}+3\alpha
_{3}\left(2n-19\right) +\frac{r_{c}^{5}+6\alpha_2r_{c}^{3}+15\alpha_3r_{c}}{\sinh r_c \cosh r_c}=0.  \label{rchyp}
\end{eqnarray}
Again for the case that Eqs. (\ref{rchyp}) have no real root, the functions $\rho $ and $\rho+p$ are
positive everywhere.
\subsection{Positivity of $\rho-\tau$}
Now, we investigate the conditions of the positivity of $\rho-\tau$.
Since $b-rb^{\prime }>0$, as one may see from Eq. (\ref{Cond1}), the
positivity of $\rho -\tau $\ reduces to
\begin{equation}
1+\frac{2\alpha _{2}b}{r^{3}}+\frac{3\alpha _{3}b^{2}}{r^{6}}<0.
\label{Exo2}
\end{equation}
One may note that when
the Lovelock coefficients are positive, the condition (\ref{Exo2}) does not
satisfy. For the cases that either of $\alpha _{2}$ and $\alpha _{3}$ or
both of them are negative, the condition (\ref{Exo2}) is satisfied in the vicinity
of the throat for power law, logarithmic and hyperbolic shape function provided that the
throat radius is chosen in the range $r_{-}<r_{0}<r_{+}$, where
\begin{equation}
r_{-}=\left( -\alpha _{2}-\sqrt{\alpha _{2}^{2}-3\alpha _{3}}\right) ^{1/2},%
\text{ \ \ }r_{+}=\left( -\alpha _{2}+\sqrt{\alpha _{2}^{2}-3\alpha _{3}}%
\right) ^{1/2}.  \label{r+}
\end{equation}
For the choices of Lovelock coefficients where $r_{+}$\ is not real, then the condition
(\ref{Exo2}) does not hold. For the cases where $r_{-}$ is not real, then there
is no lower limit for the throat radius that satisfies the condition (\ref
{Exo2}). Even for the cases where $r_+$ exists and $r_{0}$ is chosen in the range $r_{-}<r_{0}<r_{+}$,
the condition (\ref{Exo2}) will be satisfied in the region $r_{\min}<r<r_{\max}$,
where $r_{\min}$ and $r_{\max}$ are the positive real roots of the following equation:
\begin{equation}
r^6+2\alpha_2 r^3 b(r) +3\alpha_3 b^2(r)=0.  \label{rmax}
\end{equation}
For negative $\alpha_3$, Eq. (\ref{rmax}) has only one real root
and the condition (\ref{Exo2}) is satisfied in the range $0 \leq r<r_{\max}$.
It is worth noting that the value of $r_{\max}$ depends on the Lovelock
coefficients and the shape function.
The value of $r_{\max}$ for the power law shape function is
\begin{equation}
r_{\max }=\left( \frac{r_{+}}{r_{0}}\right) ^{2/(m+2)}r_{0},  \label{rmaxp}
\end{equation}
which means that one cannot have a wormhole with normal matter everywhere. It
is worth noting that $r_{+}>r_{0}$, and therefore $r_{\max }>r_{0}$, as it
should be.
\subsection{Normal and exotic matter}
\begin{figure}
\centering {\includegraphics[width=7cm]{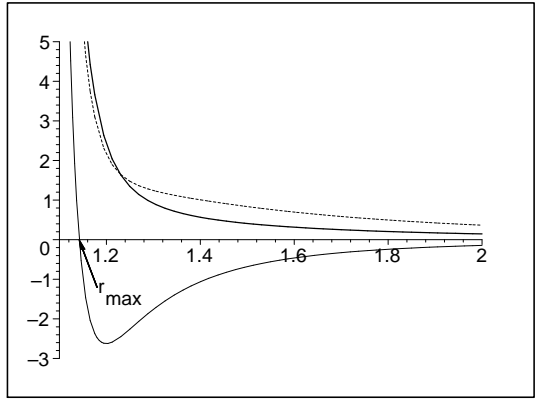} }
\caption{$\rho-\tau$ (solid line), $\rho+p$ (bold-line) and $\rho$ (dotted-line) vs. $r$  for
logarithmic shape function with $n=8$, $r_0=1.1$, $\alpha_2=-.5$, and $\alpha_3=-.5$.} \label{Log}
\end{figure}

\begin{figure}
\centering {\includegraphics[width=7cm]{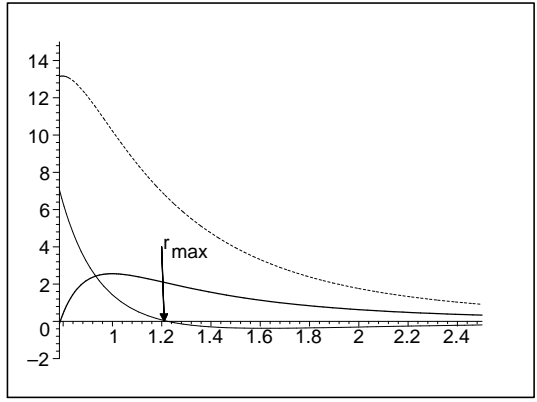} }
\caption{$\rho-\tau$ (solid-line), $\rho+p$ (bold line) and $\rho$ (dotted line) and  vs $r$  for
hyperbolic shape function with $n=8$, $r_0=r_c=.785$, $\alpha_2=-.5$, and $\alpha_3=-.5$.} \label{Hyp}
\end{figure}

\begin{figure}
\centering {\includegraphics[width=7cm]{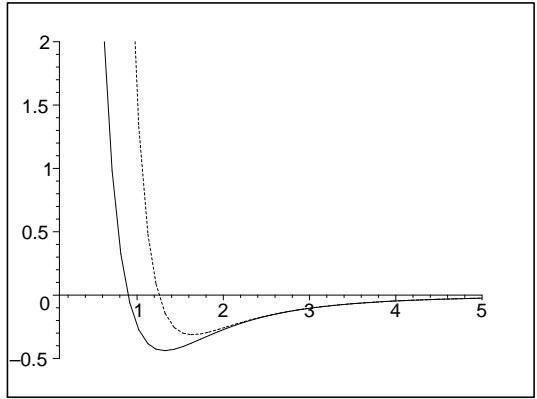} }
\caption{$\rho-\tau$ vs $r$  for
hyperbolic shape function with $n=8$, $r_0=.1$, $\alpha_2=-.5$, $\alpha_3=0$ (solid line)
and $\alpha_3=-1$ (dotted line).} \label{max}
\end{figure}

Now, we are ready to give some comments on the exoticity or
normality of the matter. First, we investigate the condition of having normal
matter near the throat. There exist two constraint on the value of $r_0$ for the power
law, logarithmic and hyperbolic shape functions, while for the logarithmic shape function
$r_0$ should also be larger than $1$. The first constraint comes from the positivity
of $\rho$ and $\rho+p$, which state that $r_0$ should be larger or equal to $r_c$,
where $r_c$ is the largest real root of
Eqs. (\ref{rcpow}), (\ref{rclog}) and (\ref{rchyp}) for power law, logarithmic
and hyperbolic shape functions,
respectively. Of course, if there exists no real root
for these equations, then there is no lower limit for $r_0$.
The second constraint, which come from the condition (\ref{Exo2}),
states that $r_+$ should be real. For positive Lovelock coefficients, there exists no real value
for $r_+$, and therefore we consider
the cases where either of $\alpha _{2}$ and $\alpha _{3}$ or both of them are negative.
The condition (\ref{Exo2}) is satisfied near the throat for the following two cases:\\
1) $\alpha_2<0$ and $0<\alpha_3 \leq {\alpha_2}^2/3$.\\
2) $\alpha_3<0$. \\
The root $r_-$ is real for the first case, while it is not real for
the second case.
In these two cases, one has normal matter in the vicinity of the throat provided
$r_c<r_+$, and $r_0$ is chosen in the range $r_{>} \leq r_0<r_+$, where $r_>$ is the largest values of
$r_c$ and $r_-$.
\begin{figure}
\centering {\includegraphics[width=7cm]{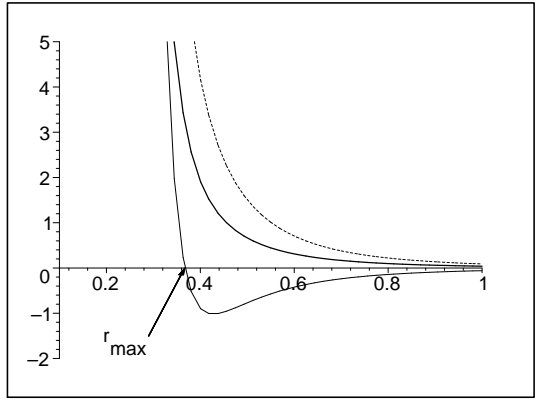} }
\caption{$\rho-\tau$ (solid line), $\rho+p$ (bold-line) and $\rho$ (dotted line) vs $r$  for
power law shape function with $n=8$, $m=2$, $r_0=.1$, $\alpha_2=-.5$, and $\alpha_3=-.5$.} \label{Pow1}
\end{figure}
\begin{figure}
\centering {\includegraphics[width=7cm]{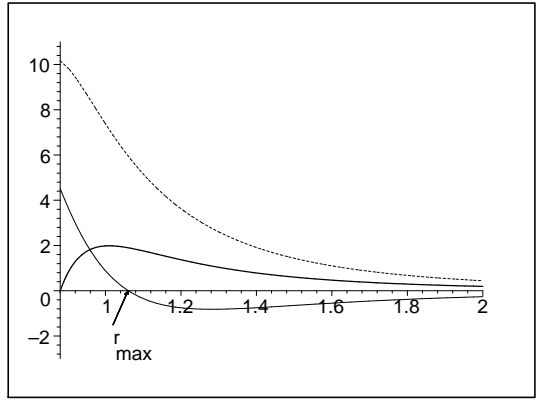}}
\caption{$\rho-\tau$ (solid-line), $\rho+p$ (bold-line) and $\rho$ (dotted-line) vs. $r$  for
power law shape function with $n=8$, $m=2$, $r_0=r_c=.88$, $\alpha_2=-1$, and $\alpha_3=.2$.} \label{Pow2}
\end{figure}
The above discussions show that there are some constraint on the
Lovelock coefficients and the parameters of shape function.
In spite of these constraint, one can choose the parameters suitable
to have normal matter near the throat. Even if the conditions of
having normal matter near the throat are satisfied, there exist an
upper limit for the radius of region of normal matter given in Eq. (\ref{rmax}).
Figures \ref{Pow1}-\ref{Hyp} are the diagrams of $\rho$, $\rho+p$ and $\rho-\tau$
versus $r$ for various shape functions. In Figs. \ref{Pow2} and \ref{Hyp},
the parameters have been chosen such that $r_c$ is real, while $r_c$ is not real
in Figs. \ref{Pow1} and \ref{Log} and therefore $r_0$
has no lower limit. Note that for logarithmic shape function
$r_0$ is larger than $1$. All of these figures show that one is able to choose
suitable values for the metric parameters in order to have
normal matter near the throat.
Also, it is worth to mention that the radius of normal matter increases as
$\alpha _{3}$ becomes more negative, as one may note from Eq. (\ref{rmax})
or Fig. \ref{max}. That is, the
third order Lovelock term with negative coupling constant enlarges the
region of normal matter.

Second, we consider the conditions where the matter is exotic
for $r\geq r_0$ with positive $\rho$ and $\rho+p$. These
functions are positive for $r>r_{0}$ provided $r_{0}>r_{c}$, where $r_{c}$
is the largest real root of Eqs. (\ref{rcpow}), (\ref{rclog}) and (\ref{rchyp}) for power
law, logarithmic and hyperbolic solutions, respectively. Of course,
there is no lower limit on $r_0$, when these equations have no real root.
On the other hand,
the condition (\ref{Exo2}) does not hold for $r_{0}>r_{+}$. Thus,
if both of $r_c$ and $r_+$ are real, and one choose $r_0 \geq r_>$,
where $r_>$ is the largest value of $r_c$ and $r_+$, then the matter
is exotic with positive $\rho$ and $\rho+p$ in the range $r_0 \leq r <\infty$.
If none of $r_c$ and $r_+$ are real, then there is no lower limit
for $r_0$, and one can have wormhole with exotic matter everywhere.

\section{Closing Remarks}

For wormholes with small throat radius, the curvature near the throat is
very large, and therefore higher order curvature corrections are invited to
the investigation of the wormholes. Thus, we presented the
wormhole solutions of third order Lovelock gravity. Here, it is worth
comparing the distinguishing features of wormholes of third order Lovelock
gravity with those of Gauss-Bonnet and Einstein gravities. While the
positivity of $\rho$ and $\rho+p$ does not impose any lower limit on $r_{0}$ in
Einstein gravity, there may exist a lower limit on the throat radius
in Lovelock gravity, which is the
largest real root of Eqs. (\ref{rcpow}), (\ref{rclog}) and (\ref{rchyp}) for
power law, logarithmic and hyperbolic shape functions, respectively. Although the
existence of normal matter near the throat is a common feature of the
wormholes of Gauss-Bonnet and third order Lovelock gravity, but the radius of
the region with normal matter near the throat of third order Lovelock wormholes
with negative $\alpha_3$ is larger than that of
Gauss-Bonnet wormholes. That is, the third order Lovelock term with negative
$\alpha _{3}$ enlarges the radius of the region of normal matter. Thus, one
may conclude that inviting higher order Lovelock term with negative
coupling constants into the gravitational field equation, enlarges the region of normal matter near the
throat. For $n$th order Lovelock gravity with a suitable
definition of $\alpha_p$ in terms of Lovelock coefficients,
the condition (\ref{Exo2}) may be generalized to
\begin{equation*}
1+\sum_{p=2}^{[n-1]/2}p\alpha _{p}\left( \frac{b}{r^{3}}\right) ^{p-1}<0,
\end{equation*}
which can be satisfied only up to a radius $r_{\max }<\infty $. One
may conclude from the above equation that as more Lovelock terms with negative
Lovelock coefficients contribute to the
field equation, the value of $r_{\max}$ increases, but one
cannot have wormhole in Lovelock gravity with normal matter everywhere for the metric (\ref{met1})
with $\phi(r)=0$. The case of arbitrary $\phi(r)$ needs further investigation.

\textbf{Acknowledgements}

This work has been supported by Research Institute for Astrophysics and
Astronomy of Maragha.

\end{document}